\documentclass{llncs}

\usepackage{listings}
\usepackage[usenames,dvipsnames,svgnames,table]{xcolor}
\usepackage{graphicx}
\usepackage{subcaption}
\usepackage{enumitem}
\usepackage{amsmath}
\usepackage{amssymb}
\usepackage{booktabs}
\usepackage{varwidth}
\usepackage{colortbl}

\definecolor{darkred}{rgb}{0.44,0,0}
\definecolor{darkgreen}{rgb}{0,0.44,0}
\definecolor{darkblue}{rgb}{0,0,0.44}
\definecolor{grey}{rgb}{0.5,0.5,0.5}

\definecolor{mygreen}{rgb}{0,0.6,0} \definecolor{mygray}{rgb}{0.5,0.5,0.5} \definecolor{mymauve}{rgb}{0.58,0,0.82}

\definecolor{dodgerblue}{rgb} {0.4,0.4,1.0} 
\definecolor{forestgreen}{rgb}{1.0,0.0,0.0} 
\definecolor{yellow1}{rgb}    {0.0,1.0,0.0} 
\definecolor{red}{rgb}        {1.0,0.0,1.0} 
\definecolor{lightgray}{rgb}{0.950,0.950,0.950}

\newcommand{\bujaruelo}{{\sc bujaruelo}}
\newcommand{\odroid}{{\sc odroid}}

\newcommand{\gemm}{{\sc gemm}}
\newcommand{\chol}{{\sc chol}}
\newcommand{\syrk}{{\sc syrk}}
\newcommand{\trsm}{{\sc trsm}}

\title{HeSP: a simulation framework for solving the task scheduling-partitioning problem on heterogeneous architectures}
\titlerunning{Titulo corto}

\author{Ant\'on Rey \and Francisco D. Igual \and Manuel Prieto-Mat\'ias}
\institute{Dept. Arquitectura de Computadores y Autom\'atica\\
Universidad Complutense de Madrid (Spain) \\
\email{\{anrey,figual,mpmatias\}@ucm.es}}

\begin{document}

\maketitle

\begin{abstract}
%
%
%
In this paper we describe HeSP, a complete simulation framework to study a general task scheduling-partitioning problem
on heterogeneous architectures, which treats recursive task partitioning and scheduling decisions on equal footing. 
Considering recursive partitioning as an additional degree of freedom, tasks can be dynamically partitioned or merged at 
runtime for each available processor type, exposing additional or reduced degrees of parallelism as needed. 
Our simulations reveal that, for a specific class of dense linear algebra algorithms taken as a driving example, 
simultaneous decisions on task scheduling and partitioning yield significant performance gains 
on two different heterogeneous platforms: a highly heterogeneous CPU-GPU system and a low-power asymmetric 
big.LITTLE ARM platform.
The insights extracted from the framework can be further applied to actual runtime task schedulers in order
to improve performance on current or future architectures and for different task-parallel codes.

\end{abstract}

\section{Introduction and motivation}


%
Task-parallel programming models have emerged as an appealing solution in 
order to tackle the programmability problem on both homogeneous and heterogeneous platforms.
These efforts aim at reducing user intervention to manage data dependences, task allocation and data
transfer management by delegating those tasks to underlying runtime task schedulers.
%
%
However, the ever-increasing heterogeneity in current (and future) architectures has dramatically aggravated the challenge for
runtime developers; as more types of computing resources are available, it becomes more difficult 
to concurrently exploit them in order to optimize co-operative parallel implementations. 
One of the main conceptual problems lies  
on how to optimally (and possibly dynamically) partition a task into sub-tasks (that is, solving a {\em task
partitioning problem}), and how to efficiently schedule them 
to the most convenient resource among those available in order to maximize performance 
(that is, solving a {\em task scheduling problem}).

In this paper, we present HeSP ({\em Heterogeneous Scheduler-Partitioner}), a simulation 
framework that addresses both problems in a simultaneous fashion. Based on per-task and data transfers performance 
models, HeSP adds an additional degree of freedom to typical task scheduling policies by considering a 
joint task partitioning/scheduling approach.  The framework proceeds by 
finding a set of task partitions that divides 
the initial workload into a number of sub-tasks with different granularity, that best fit to the underlying hardware 
resources at a given execution point. The approach drives to considerable performance improvements and more efficient resource utilization. We show that 
the new task scheduler-partitioner paradigm is of wide appeal to increase the scheduling quality on highly 
heterogeneous architectures, and to gain insights that can be further applied to specific task-parallel implementations,
actual runtime task schedulers, and present and future heterogeneous architectures.



Runtime task schedulers are capable of managing efficient load balancing, asynchronous out-of-order task execution and handling data 
across separated memory spaces, abstracting these mechanisms
to the programmer.
%
Concretely, StarPU~\cite{StarPU}, OmpSs~\cite{OmpSs} or XKaapi~\cite{XKaapi}, among others, offer implicit 
parallel programming models with transparent data dependence analysis among tasks, and support scheduling on heterogeneous processing platforms. 
Efficient scheduling under this task-based perspective strongly depends on the quality of the scheduling policies implemented in the runtime,
and more specifically, how they address the special features of the algorithm and the underlying architecture. 

These efforts usually consider the static creation and management of equally-sized tasks operating
on uniform data tiles, which naturally drives to an improper load balancing 
among computing resources on heterogeneous architectures, given the different processing capabilities of each type of resource. As a side effect,
the optimal block size, even in the homogeneous case, is a time-consuming effort for the developer, and strongly depends on the algorithmic properties
of the target implementation and the specific features of the underlying architecture.
Although each processor type typically reaches its performance peak for substantially different task sizes, and the chosen initial granularity exposes a fixed amount of parallelism, 
few strategies have been developed in order to dynamically adapt task granularity to the underlying heterogeneous hardware.
%
Focusing on dense linear algebra implementations,~\cite{BosilcaHierDAG} proposes a hierarchical directed acyclic graph (DAG onwards) strategy, creating a two-level DAG hierarchy on systems featuring 
two types of computing platforms (CPU/GPU). Similarly,~\cite{Haidar} proposes an offline adaptation of the task grain size to the processor type and
to statically assign tasks to distributed compute nodes. On the other hand,~\cite{Cojean} proposes an alternative approach in which computing resources are aggregated as needed in order to adapt the computing capabilities to coarse grain kernels.
The {\em Versioning} task scheduler for the OmpSs runtime~\cite{versioning} defines multiple implementations per task,
each one targeting a different processor type, and decides at runtime where to map them based on historical runtime information. 

HeSP extends the aforementioned efforts by exploring the global impact of {\em arbitrary} 
degrees of task granularity on an arbitrary heterogeneous platform, adapting task sizes not only to the individual processor capabilities, but also to the current degree of available parallelism dictated by a specific algorithm.  

\subsection{A motivating example: tiled Cholesky factorization}

Let us expose a motivating and illustrative example of the actual problems
related with equally-sized task partitioning on heterogeneous platforms. 
%
The blocked Cholesky factorization decomposes an $n \times n$ symmetric positive definite matrix
$A$ stored by $s \times s$ blocks of dimension $b \times b$ each, into $A=LL^T$ where $L$ is a lower triangular matrix. At runtime, the outer
loop in the code depicted in Figure~\ref{lst:chol} that calculates the Cholesky 
factorization, divides the operation into a number of sub-tasks that, when executed under a task-parallel
paradigm, generate a task DAG as that shown in Figure~\ref{fig:dag}. In the task DAG, 
nodes correspond to different tasks, and edges denote data dependencies between them.


\begin{figure}[t]
\begin{lstlisting}
void cholesky (double *A[s][s], int b, int s) {
  for (int k = 0; k < s; k++) {
     chol (A[k][k], b, b);              // Cholesky factor. (diag. block)

     for (int j = k + 1; j < s; j++) 
        trsm (A[k][k], A[k][j], b, b);  // Triangular solve

     for (int i = k + 1; i < s; i++) {
        for (int j = i + 1; j < s; j++)        
           gemm (A[k][i], A[k][j],      // Matrix multiplication
                 A[i][j], b, b);

        syrk (A[k][i], A[i][i], b, b);  // Symmetric rank-b update
} } }
\end{lstlisting} 
\caption{\label{lst:chol} C implementation of the blocked Cholesky factorization.}
\end{figure}

%

  \begin{figure}[t]
    \centering
    \begin{subfigure}[b]{0.48\textwidth}
        \centering
        \includegraphics[angle=90,height=2.5cm,width=5.4cm]{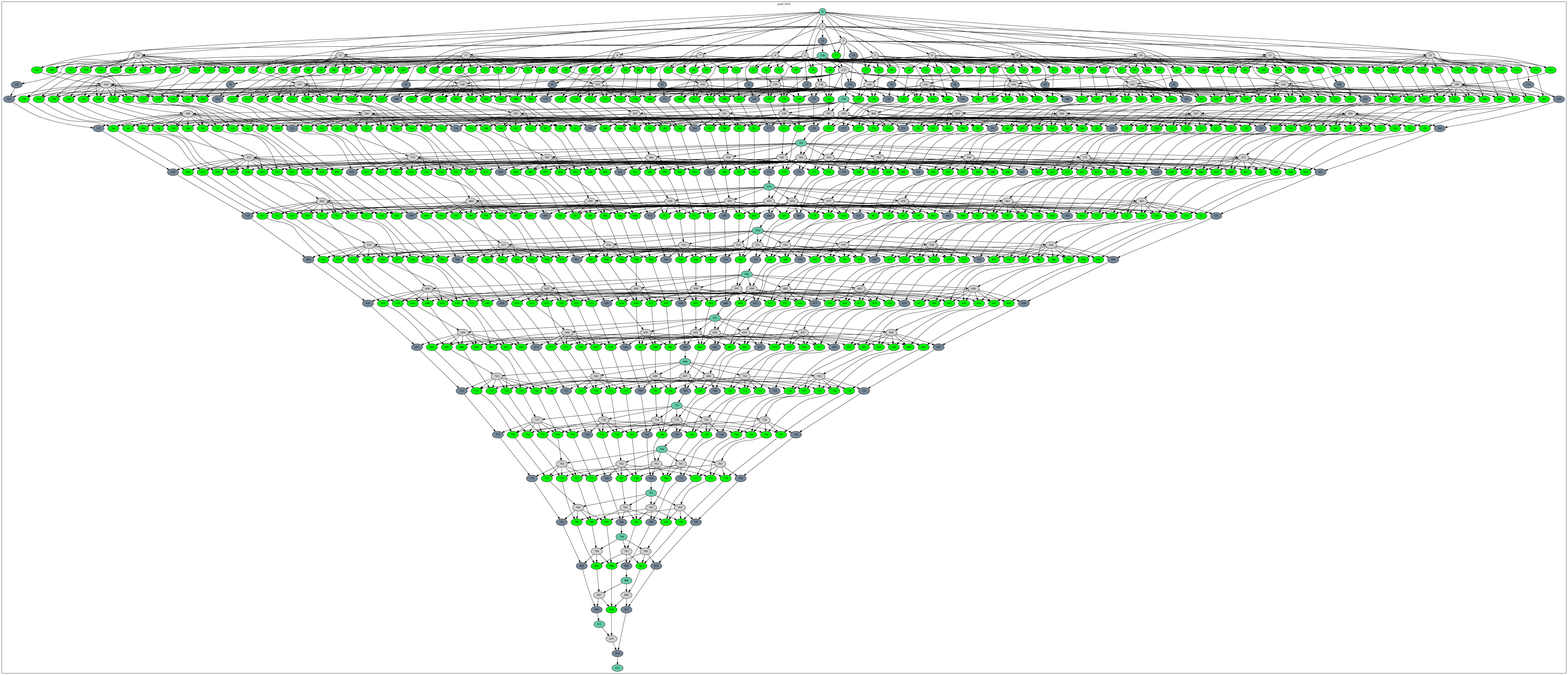}
        \caption{Task DAG.}
        \label{fig:dag}
    \end{subfigure}
      ~
    \begin{subfigure}[b]{0.48\textwidth}
        \centering
        \includegraphics[height=2.5cm,width=6.0cm]{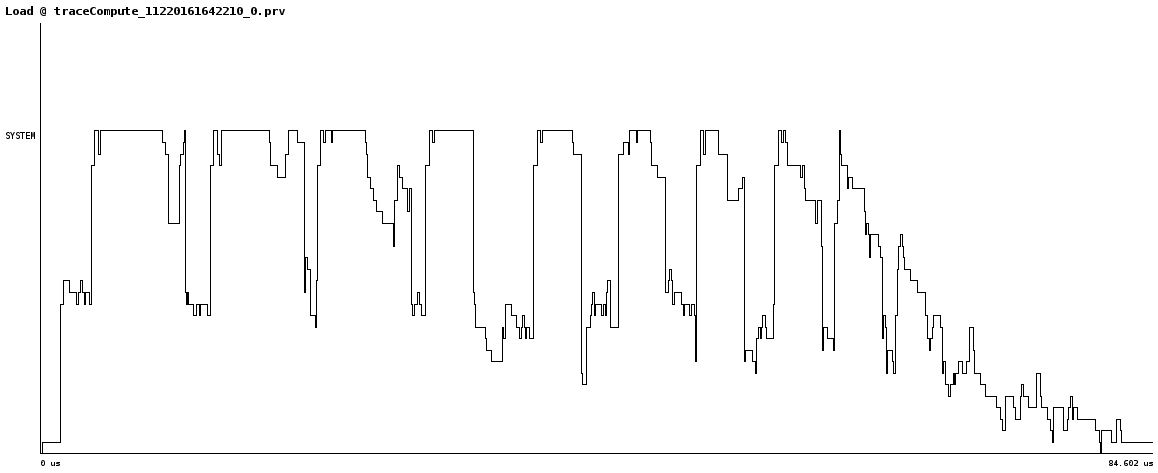}
        \caption{Compute load trace.}
        \label{fig:load}
    \end{subfigure}
\caption{
\label{fig:dag_load}%
(a) Task DAG in which the computation evolves from left to right, and (b) compute load trace generated by the Cholesky factorization in Figure~\ref{lst:chol}, for a problem size $n=16,384$, and block size $b=1,024$.}
\end{figure}

The Cholesky factorization is an appealing example for our purposes: 
it exhibits different sub-task types (\chol, \syrk, \gemm\ and \trsm) and complex
data dependences among them, and it features different degrees of parallelism as
the factorization evolves. Consider, for example, how the DAG depicted in Figure~\ref{fig:dag} 
reduces the potential parallelism (that is, the number of tasks that can be potentially executed 
in parallel, typically related with the width of the DAG) at the first stages of the factorization, and (in a much larger extent) at the last 
stages.
This is usually translated into processor load patterns like that shown in Figure~\ref{fig:load},
that represents a timeline of an execution of the Cholesky factorization on a highly heterogeneous platform, composed by 28 Intel Xeon cores and 3 different GPUs. 
The plot represents the number of active processors as the execution proceeds. Areas with reduced load are usually 
due to load imbalance. Note that this phenomenon can be motivated by two different
factors: different processing capabilities of each processor type, and lack of potential parallelism on specific stages
of the execution. The first can be alleviated by {\em scheduling} heuristics (e.g. mapping tasks in the critical path to fast
processors), but the second is inherent to the algorithm, and can be alleviated by dynamic task {\em partitioning} in order
to expose additional parallelism at runtime.

Data block (tile) size is a crucial parameter in task-parallel executions, as it ultimately
determines the amount of available parallelism, and the efficiency of each individual task
execution. In Figure~\ref{lst:chol}, block size is determined by {\tt b}; note that,
typically, larger block sizes usually imply higher performance per individual task,
and smaller block sizes tend to expose higher degrees of parallelism, which naturally drives to
better processor occupation. In addition, different block (task) sizes
are desired for different architectures, and even for different problem dimensions
in the same architecture. 

Altogether, these observations motivate the exploration of new techniques that explore
the impact of \textit{heterogeneous} or \textit{non-uniform} task partitioning on the performance and resource occupation
of heterogeneous architectures.  
In the following, we introduce HeSP, a complete framework that supports the definition
of complex heterogeneous architectures and simulates simultaneous task-scheduling and
task-partitioning schemes that alleviate the aforementioned problems.

\section{HeSP: Heterogeneous Scheduler-Partitioner}

HeSP is a simulation framework that statically explores
the task scheduling-partitioning problem targeting heterogeneous architectures.
At a glance, the input to this problem is {\em (1)} a hardware platform description where several finite-size memory spaces 
are connected according to a certain network topology, together with a (possibly heterogeneous) set of processors associated with them; and {\em (2)}, a task to be computed in that platform.
A solution to this problem consists of {\em (1)} a set of tasks --presumably with different granularity--, related by arbitrary data dependences and equivalent to the input task, and {\em (2)} a task-to-processor mapping.
The objective function is typically performance maximization, although energy consumption minimization is also supported by HeSP.

\subsection{Features of the scheduling-partitioning simulation framework}

Besides supporting recursive task partitioning, HeSP is designed to be a realistic framework that simulates 
not only current heterogeneous architectures, but also state-of-the-art scheduling and data management policies on task-parallel executions. In the following, we introduce its features in detail.

\vspace{0.1cm}
{\bf Task and data scheduling heuristics.}
%
HeSP implements different heuristics for task-to-processor assignments.
\textit{Random} ({\sc R-P}) and \textit{Fastest} ({\sc F-P}) processor selection policies consider such processor choices among idle processors at the task release time.
The \textit{Earliest Idle Time} ({\sc EIT-P}) and \textit{Earliest Finish Time} ({\sc EFT-P}) policies select the processor becoming idle first, and the processor finishing first if that task is assigned to it, respectively. {\sc EFT-P} estimates the finishing time accounting for eventual data transfers if needed.
Task scheduling order is specified by choosing between {\em First-come, first-served} ({\sc FCFS}) or {\em Priority-List} ({PL}) choices.
In {\sc PL}, a priority list is built by sorting tasks by their critical times in decreasing order. Critical times are computed by averaging task processing time for all processors, and propagating them throughout the task DAG by a backflow algorithm.
The combination of {\em Priority-List} and {\sc EFT-P} heuristics is practically identical to the well-known \textit{HEFT} scheduling algorithm~\cite{HEFT}.

%
%
%
%
%


When several independent memory spaces are present, HeSP considers data movement for scheduling decisions, considering individual memory spaces of each accelerator as software caches of a main memory space, typically tied to CPUs.
Common caching policies like {\em write-through} ({\sc WT}), {\em write-back} ({\sc WB}) or {\em write-around} ({\sc WA}) are used.
%
When a task is about to be scheduled to a processor, the required data transfers are issued from the source memory space to the memory space the processor is tied to using prefetching schemes. 

\vspace{0.1cm}
{\bf Performance and data transfer models.}
HeSP estimates computing or transfer times relying on performance and data transfer models extracted a priori
for each existing processor/interconnect in the system. These estimations are required when making both scheduling and partitioning decisions, jointly or in an isolated fashion.
The quality and accuracy of performance models will ultimately determine the accuracy of the simulated scheduling results. 

\begin{figure}[ht]
\begin{multicols}{5}
  \centering
\null \vfill
  {\includegraphics[height=0.55in]{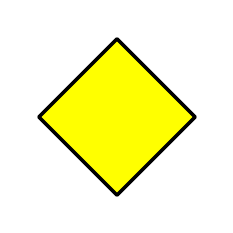}}
\vfill \null
\columnbreak
\null \vfill
  {\includegraphics[height=1.4in]{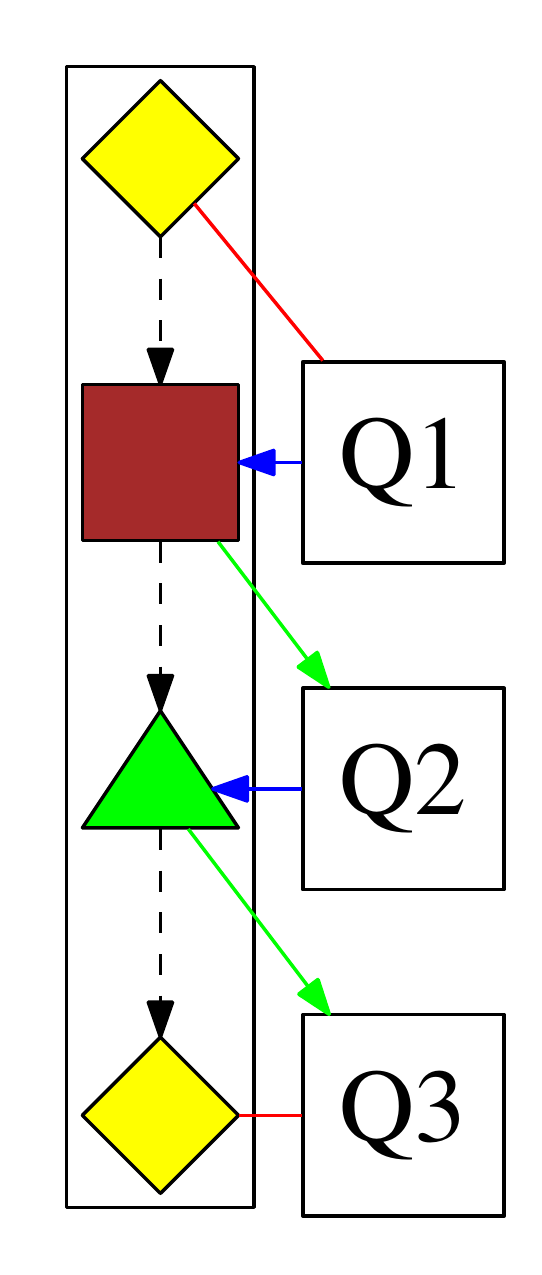}}
\vfill \null
\columnbreak
\null \vfill
  {\includegraphics[height=1.4in]{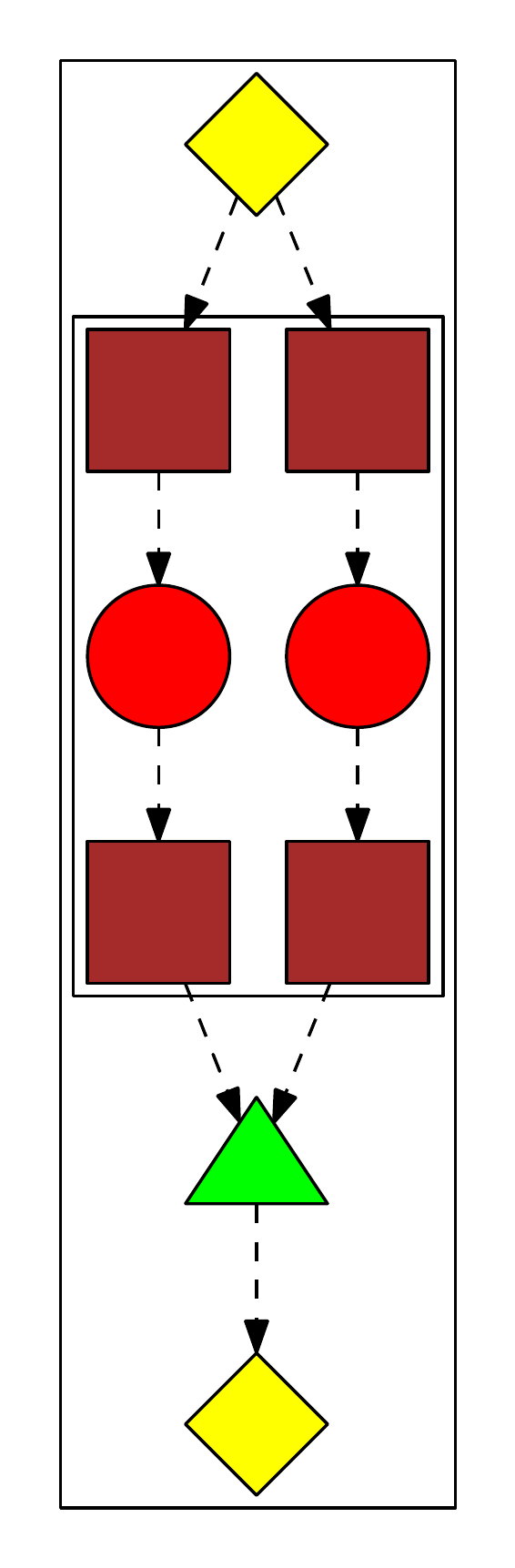}}
\vfill \null
\columnbreak
\null \vfill
  {\includegraphics[height=1.4in]{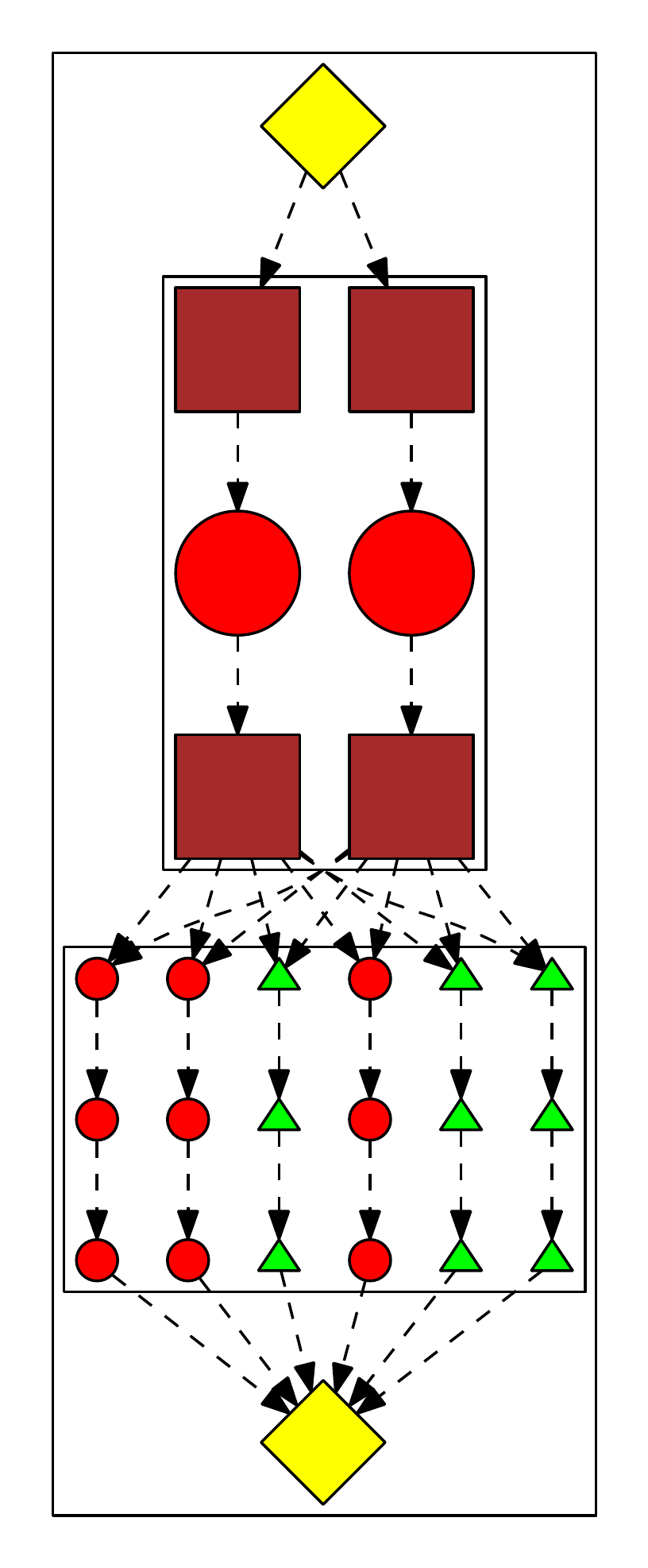}}
\vfill \null
\columnbreak
\null \vfill
  {\includegraphics[height=0.9in]{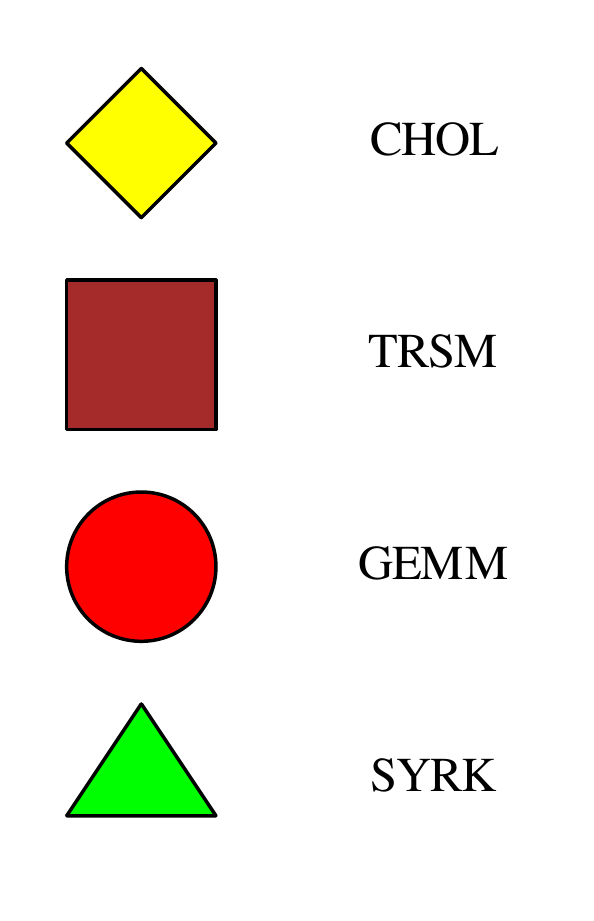}}
\vfill \null
\end{multicols}
\vspace{-0.5in}
\begin{multicols}{5}
  \centering
\null \vfill
  {\includegraphics[height=0.6in]{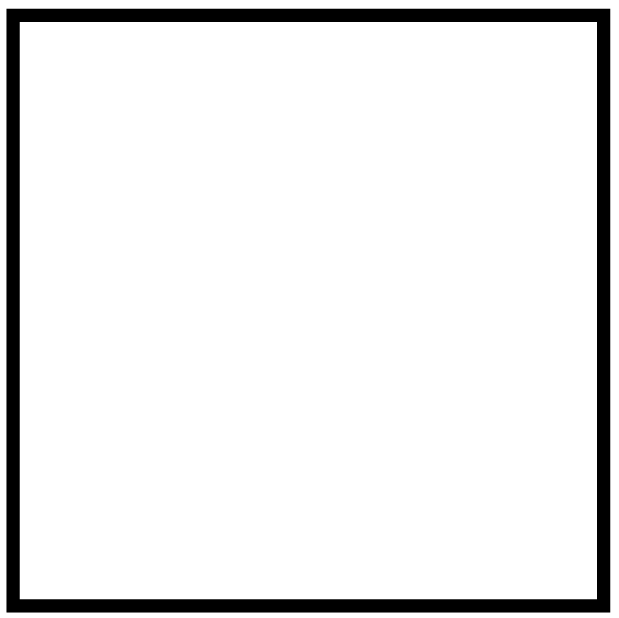}}
\vfill \null
\columnbreak
\null \vfill
  {\includegraphics[height=0.6in]{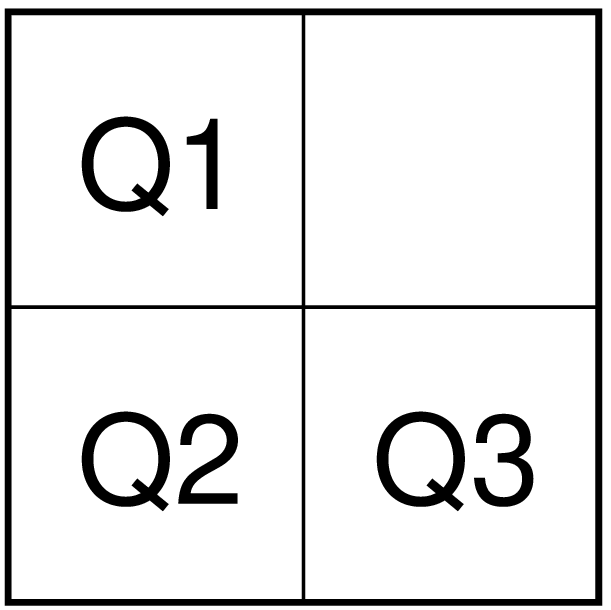}}
\vfill \null
\columnbreak
\null \vfill
  {\includegraphics[height=0.6in]{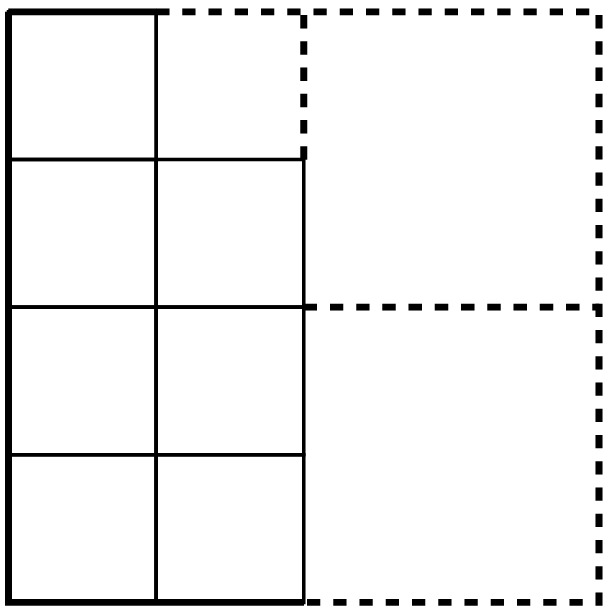}}
\vfill \null
\columnbreak
\null \vfill
  {\includegraphics[height=0.6in]{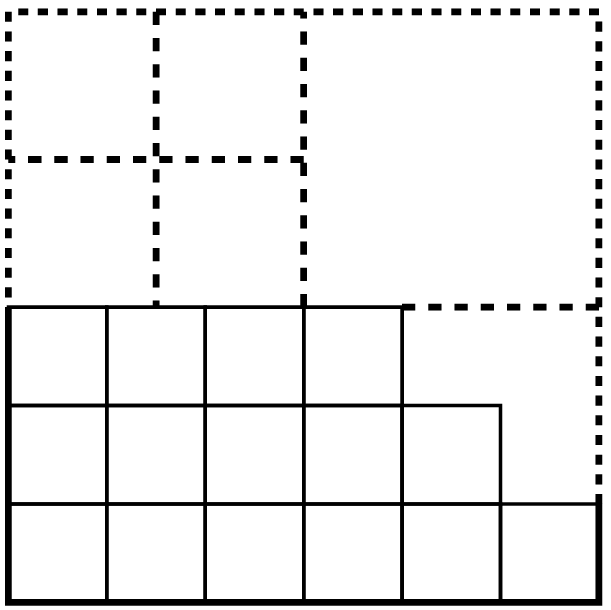}}
\vfill \null
\columnbreak
\null \vfill
  {\includegraphics[height=0.6in]{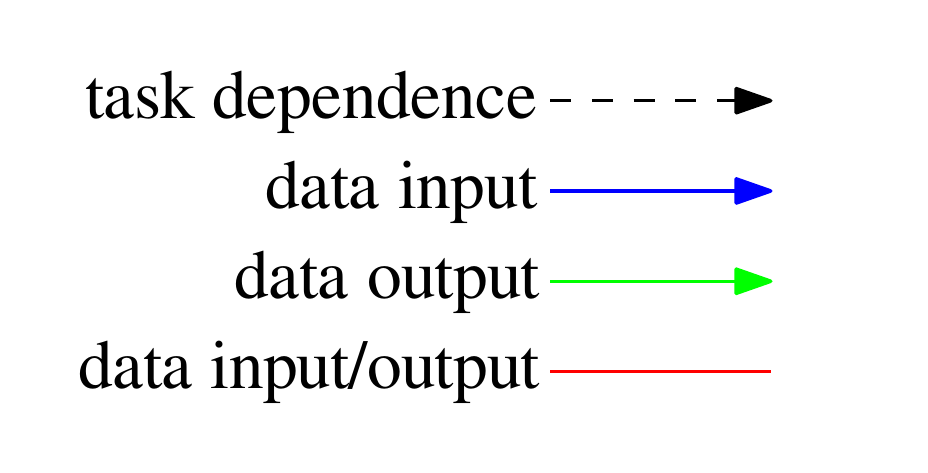}}
\vfill \null
\end{multicols}
%
  \caption{Three successive task partitions and corresponding partitioned data blocks. Tasks and their related data dependences --Q1, Q2 and Q3 quadrants-- are shown only for the first partition for the sake of clarity.}
  \label{fig:taskdata_partitions}
\end{figure}

\vspace{0.1cm}
{\bf Recursive task partitioners.}
Task \textit{partitioners}, specified for each task type willing to be partitioned, are just blocked algorithms (see, for example, Figure~\ref{lst:chol} for the specific case of the Cholesky factorization) with an input parameter that specifies the data granularity/degree of parallelism of the following partition. 
On a partitioner invocation, the corresponding emergent sub-tasks are managed by HeSP by introducing them in the respective task DAG which the partitioned task belongs to. In Figure~\ref{fig:taskdata_partitions}, starting from initial \chol\ 
task --Cholesky factorization--, it is illustrated how three successive task partitions --corresponding to respective \chol, \trsm\ and \syrk\ blocked algorithms-- affect the prior task DAG, and the corresponding data partitions they induce.

%

Note that any task can be partitioned again as long as its dependent data blocks can be divided consistently, so an extremely hierarchical task DAG can be constructed by recursively partition its tasks. A \textit{task cluster} 
is a set of tasks generated from a single task partitioning, being the source task their \textit{parent}. 
We refer to the \textit{DAG/graph depth} to indicate the maximum number of nested task clusters, and \textit{DAG/graph width} as the maximum number of tasks that can be run in parallel. 
For instance, in the four task DAGs in Figure~\ref{fig:taskdata_partitions}, the corresponding depths are 0, 1, 2 and 2, and their widths are 1, 1, 2 and 6.
Dependences between tasks, shown as dashed arrows, represent RaW, WaR and WaW constraints.

\vspace{0.1cm}
{\bf Recursive data partitioning and data coherence management.}
New tasks generated after a partition reference to finer-grained input and output data dependences, which are partitions of the initial data block(s) of the parent task (see Figure~\ref{fig:taskdata_partitions}).
HeSP implements validate/invalidate mechanisms to ensure data coherency among different memory spaces while handling 
asynchronous memory transfers. Since recursive task partitions induce corresponding recursive data block partitions, 
the existing partitioned data blocks are organized in a directed acyclic graph structure (data DAG onwards) 
in which nodes represent data blocks and directed links represent nesting relations between them;
for example, $A \longrightarrow B$ means $B$ is fully contained in $A$ and $A$ is bigger than $B$. 

Armed with the data DAG, validations and invalidations are propagated by top-bottom and bottom-top mechanisms throughout this graph to maintain coherence. For instance, to ensure that a task can start its computation and store the result in an output block $OB$, not only the block $OB$ must be invalidated on the remaining memory spaces in which the block might be present, but the hypothetical data block partitions contained in $OB$ and the bigger blocks in which $OB$ might be contained must be invalidated as well. Similarly, after a certain task has finished its computation updating $OB$, both $OB$ and all the blocks within $OB$ must be validated in the memory space corresponding to the processor assigned to that task. 

\begin{figure}[t]
\begin{multicols}{2}
 \includegraphics[width=0.35in]{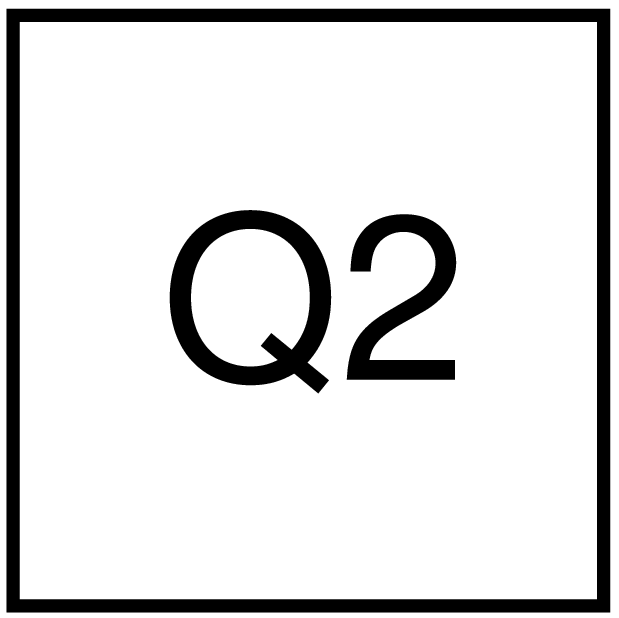}
 \includegraphics[width=0.35in]{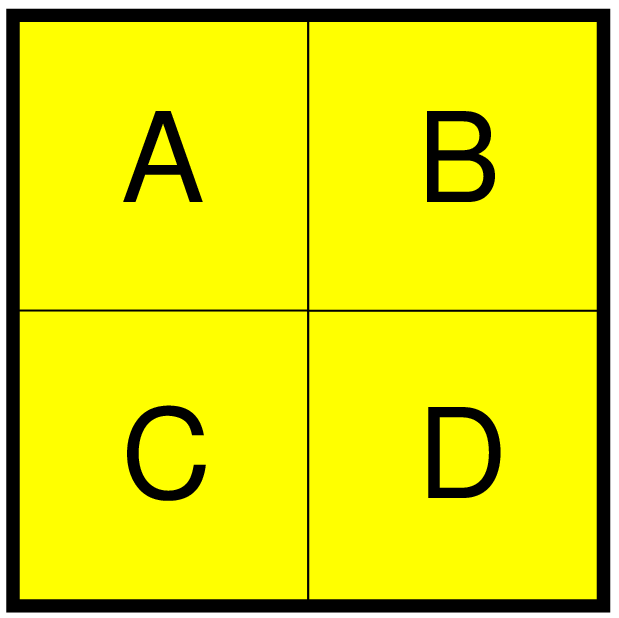}
 \includegraphics[width=0.35in]{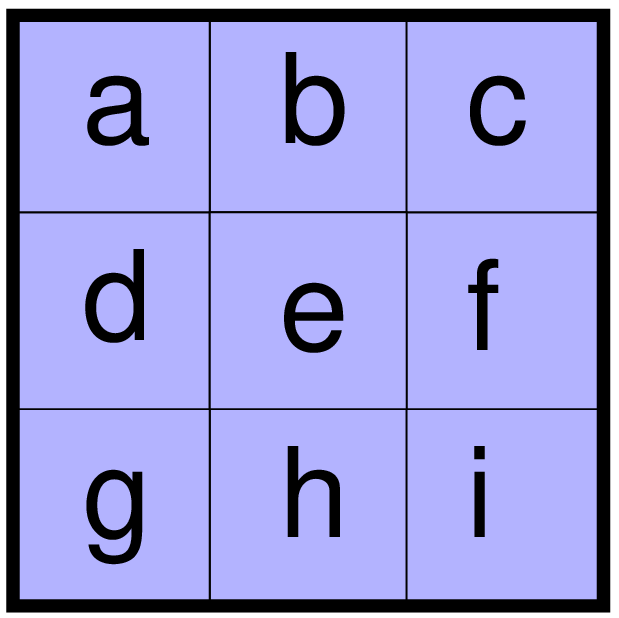}
\columnbreak
 \includegraphics[width=3.6in]{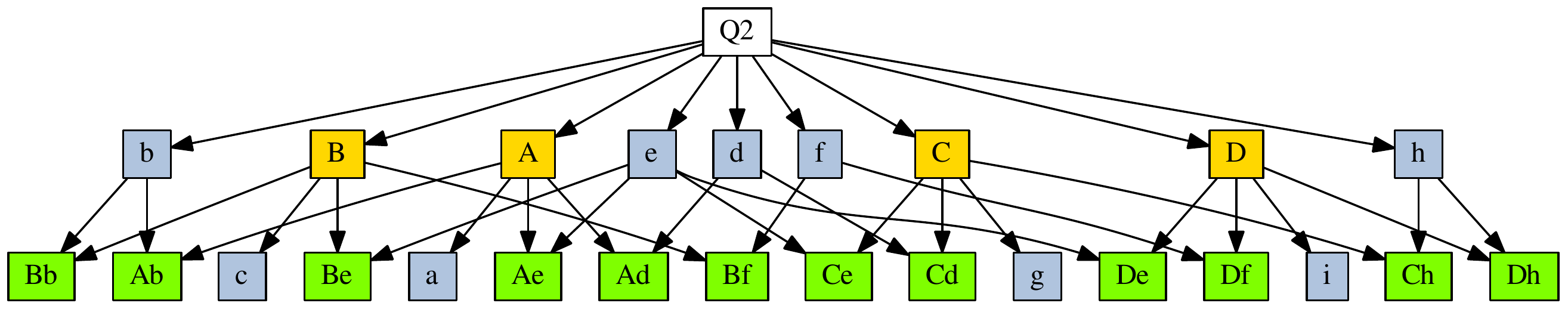}
\end{multicols}
 \caption{A data block (Q2 quadrant) can be simultaneously divided according to different tilings (yellow and blue tilings, 
corresponding to \trsm\ and \syrk\ task partitions in Figure~\ref{fig:taskdata_partitions}). Additional data block descriptors (green) are constructed to represent partial overlaps between not nested data blocks.}
 \label{fig:data_dag}
\end{figure}

In general, these data block partitions induced by task partitions form tree structures. However, it is possible to have a pair of blocks which intersect partially, nested within a common bigger parent block. This case shows up when two partitions of non-divisible grain sizes are applied to the same data block (for example, quadrant $Q2$ in Figure~\ref{fig:taskdata_partitions}).
In this case, a new data block descriptor which refers to its intersection is introduced in the data DAG as a common child node of two intersecting 
blocks (see Figure~\ref{fig:data_dag}). With this mechanism, together with the validate/invalidate propagation mechanisms, data coherency is ensured for all possible data partitions and hierarchical data graphs.

\vspace{0.1cm}
{\bf Iterative solver.}
HeSP solves the scheduling-partitioning problem by iteratively searching for those hierarchical task DAGs which 
best fit to an heterogeneous processing platform --according to performance optimization-- given a specific combination of the mentioned scheduling heuristics --processor selection heuristic and task ordering--. A schedule stage is followed by a partition stage for each iteration, being the number of iterations an user-defined parameter.

At the partition stage, HeSP chooses a candidate task to be partitioned or a candidate task cluster to be merged back/repartitioned with a different granularity. 
A global analysis of the schedule-partition done in the previous iteration can provide useful information --i.e., bottleneck identification, number of idle resources, or too fine grained tasks--
 to help the iterative process to converge towards a better overall schedule-partition. This is the reason why we chose an iterative approach as the first implementation of HeSP 
instead of a \textit{local-scoped} constructive approach, in which scheduling or partitioning decisions are made at every task arrival to the scheduling queue.


The partition procedure is based on two stages: {\em (1)}, task selection to build the candidate list, and {\em (2)}, sampling type to choose the final candidate. 
For {\em (1)}, HeSP implements three different policies: \textit{All}, \textit{CP} and \textit{Shallow}. \textit{All} selects all tasks of the previous step, 
\textit{CP} selects only tasks belonging to the critical path, and \textit{Shallow} selects those tasks whose depth (that is, number of task clusters that contain it) is minimal. 
All existing task clusters are candidates to be merged back or repartitioned. For each added candidate, a positive score is computed by subtracting the current cost 
delay by an estimated cost after its eventual partition or merge, being this estimation based on the available parallelism at its scheduling time of the previous step. For each candidate whose data dependences have a characteristic size $d$, a partition parameter $p\in (0,1]$ is chosen such that new tasks created after the eventual partition will depend on data blocks of size $b=p\times d$\footnote{A task cluster is a candidate to be merged if $p=1$ or repartitioned if $p<1$.}. The more available parallelism is exposed, the smaller $p$ is set in order to generate a higher amount of parallel finer-grained tasks.

In the second stage, a final selection among all candidates is done according to \textit{Hard} or \textit{Soft} procedures. In \textit{Hard}, the candidate with the maximum score is chosen;
in \textit{Soft}, the candidate is randomly selected such that the selection probability equals the score divided by the sum of all scores.

\section{Performance results on heterogeneous architectures}


In the following, we will feed HeSP with data describing two different heterogeneous architectures: \bujaruelo, a highly heterogeneous CPU-GPU architecture, composed by 
28 Intel Xeon-E5 2695v3 cores running at 2.3 GHz, 2 GeForce GTX980 GPU and 1 GTX950;  and \odroid, a low-power asymmetric ARM architecture
with two types of processors: 4 slow Cortex-A7 and 4 fast Cortex-A15, running at 800 and 1300 MHz respectively. 
{\sc Nvidia cuBLAS/cuSOLVER} v7.5 and Intel MKL v11.3 were
used to extract task performance models on \bujaruelo, and BLIS v0.9.1 was used on \odroid.

\subsection{Framework validation and evaluation of scheduling heuristics}

The goal of the first set of experiments is twofold: first, to validate the results extracted from HeSP by comparing them with an equivalent execution 
using a real task scheduler; second, to illustrate the impact on performance of several scheduling policies in HeSP when using {\em homogeneous} or {\em uniform} task partitions. 

\begin{figure}[t!]
 \includegraphics[width=0.5\textwidth]{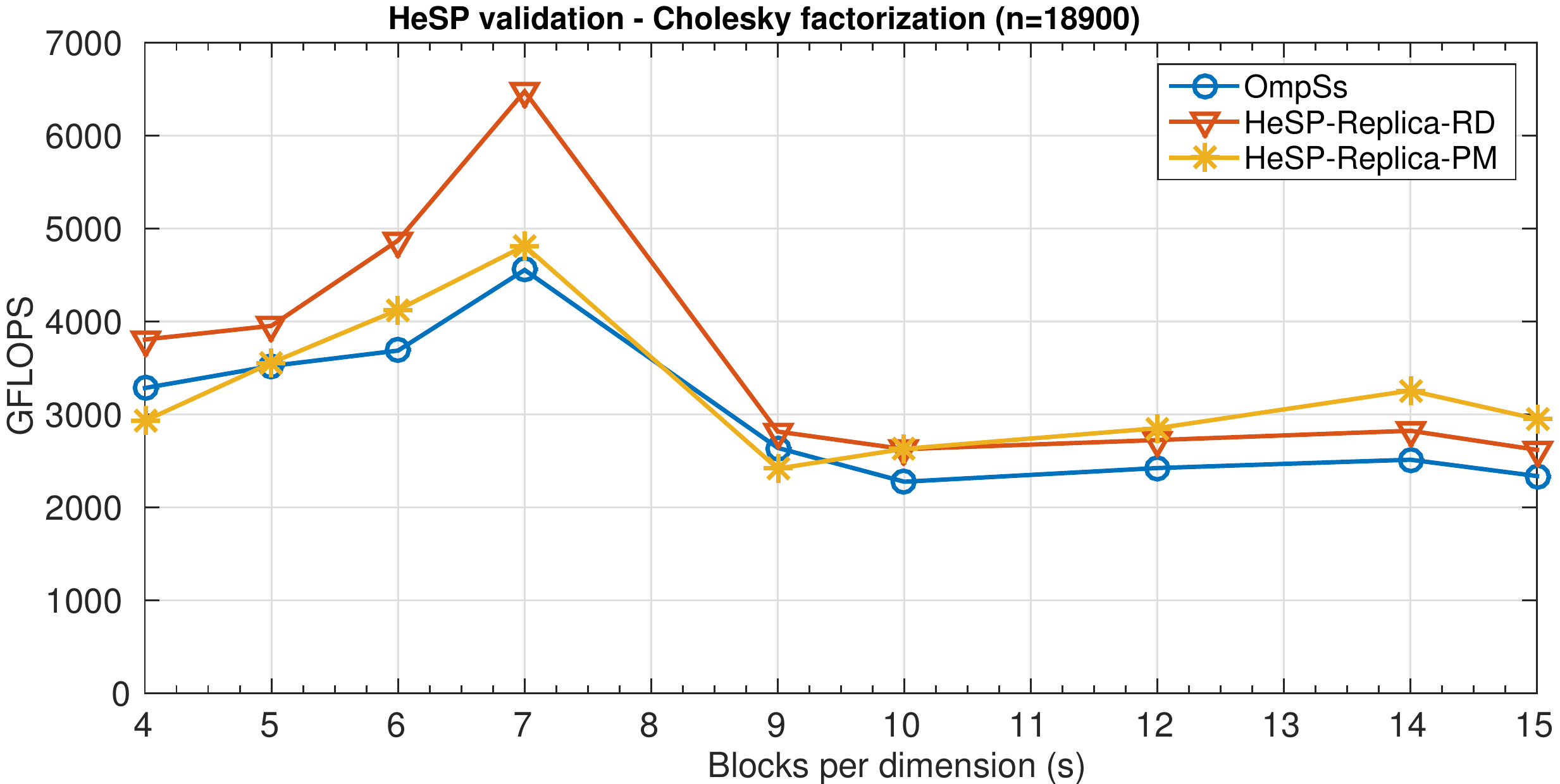}
 \includegraphics[width=0.5\textwidth]{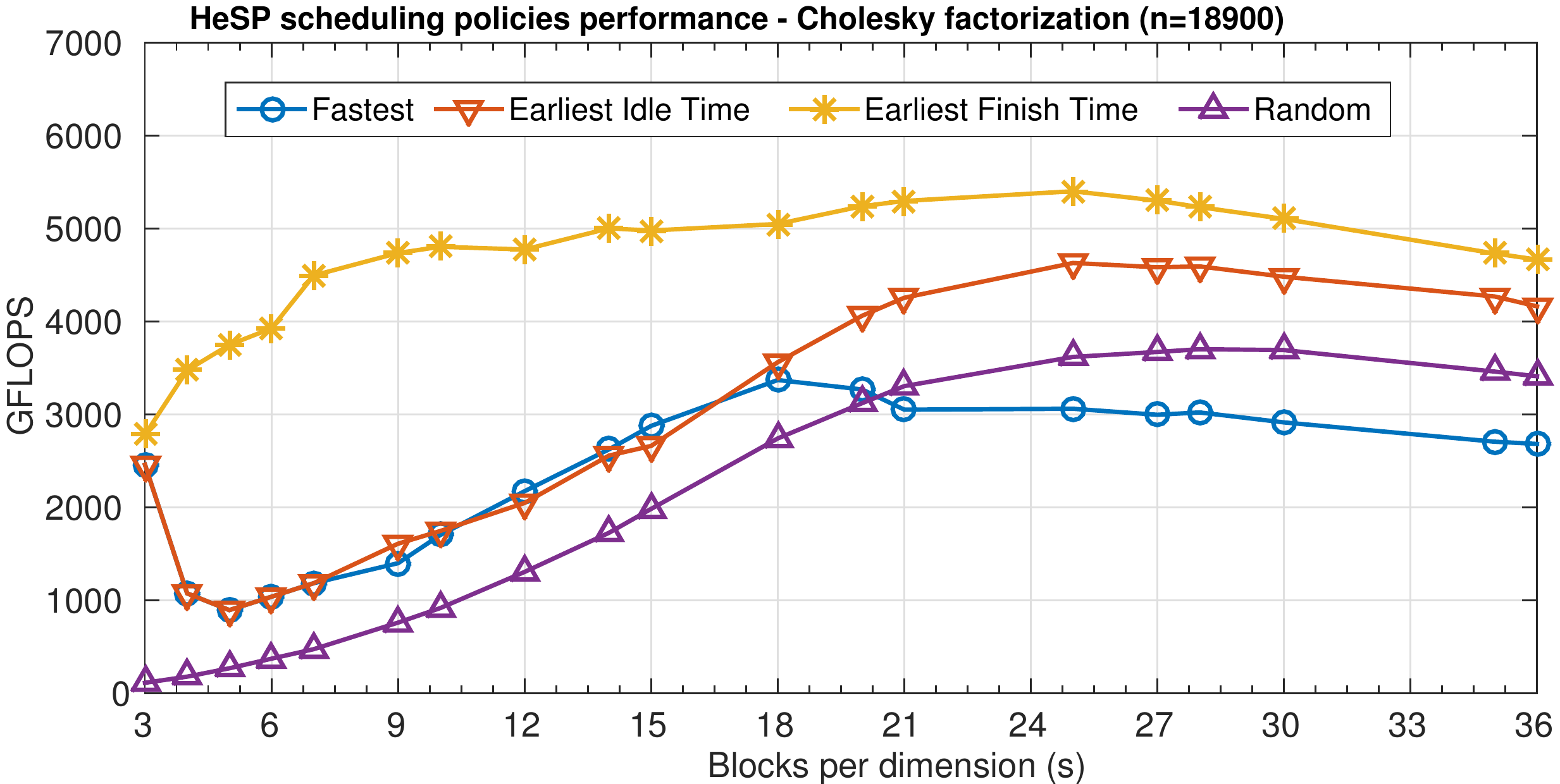}
 \caption{Left: Comparison between OmpSs and their replicated schedules. 
Right: Comparison between different scheduling policies and block sizes in HeSP.} 
 \label{fig:validation}
\end{figure}

Each point in the OmpSs line in Figure~\ref{fig:validation} (left) corresponds to the best scheduling performance out of 20 OmpSs executions for each grain size. These 20 trials were set to let OmpSs \textit{Versioning} scheduler~\cite{versioning} improve itself by gathering enough task execution delay samples for each task type/size and processor.
The other two curves --{\sc HeSP-Replica-PM} and {\sc HeSP-Replica-RD}-- denote the performance attained by HeSP when applying the same task-to-processor mapping extracted from the best OmpSs trial, using our performance models and the real OmpSs task delays, respectively, for each homogeneous tiling.

Differences in performance between {\sc HeSP-Replica-RD} and {\sc OmpSs} points are a measure of the OmpSs runtime overhead while the differences between {\sc HeSP-Replica-PM} and {\sc HeSP-Replica-RD} are mainly due to the accuracy of our performance models and possible differences between own OmpSs task delay instrumentation module and the instrumentation we used to extract our performance models.
%
Summarizing, the differences between the replicated schedules are small enough and easily explainable to assert the validity of the following results. In general,
our observations reveal a qualitative matching between real and simulated workloads for all problem sizes, with deviations that can be easily explained and do not
usually affect the quality of the observations. 

To introduce the context where our \textit{heterogeneous} or \textit{non-uniform} partitioning approach takes place and its potential benefits, 
Figure~\ref{fig:validation} (right) reports the performance obtained by running HeSP simulations using different 
scheduling policies for different number of tiles, using homogeneous task partitions.  
Some facts are remarkable: first, the optimal tile size does not only depend on the underlying architecture and problem size, 
but also on the selected scheduling policy; second, for every policy, performance curves follow a similar pattern, exhibiting a peak
performance in a trade-off tile size that best balances potential parallelism and optimal individual task performance; third, 
differences in performance are relevant depending on the selected scheduling policy, being even more dramatic for large tile sizes; 
this gives a clue on the potential benefits that will be obtained by using an heterogeneous partitioning scheme, as exposed next. 

\subsection{Impact of hetereogeneous partitioning on performance}

In the following, we illustrate the main performance improvements obtained with HeSP using \textit{All/Soft} configuration for task partitioning selection.
Table~\ref{tab:performance} reports performance values on \bujaruelo\ and \odroid\ using the best homogeneous and heterogeneous
partitions found by HeSP for different scheduling policies\footnote{In all cases, we use {\sc WB} as the caching mechanism.}, together
with additional metrics that clarify many of the concepts exposed hereafter, including average processor load, optimal/average block size and task
DAG depth. The first point to notice is the overall improvement attained for all heterogeneous task partitions found by HeSP and
the overall reduction in the optimal average block size on heterogeneous partitions.
\newcommand{\gray}{\rowcolor[gray]{.90}}

\begin{table}[t!]
\centering
	\begin{scriptsize}
\caption{Performance comparison for \bujaruelo\ and \odroid.}
\label{tab:performance}
\begin{tabular}[t]{|c||c|c|c||c|c|c|c|c|}
\hline 
\rowcolor[gray]{.9}
\multicolumn{9}{|c|}{\bujaruelo\ ($32,768\times 32,768$ Cholesky factorization in single precision)} \\ \hline \hline
 & \multicolumn{3}{c||}{\bf Best Homogeneous} & \multicolumn{5}{c|}{\bf Best Found Heterogeneous} \\ \hline \hline
    & Perf.       & Avg. load & Block & Perf.         & Improve.    & Avg. load  & Avg.      & DAG  \\
Config.  & (GFLOPS)    & (\%)      & size  & (GFLOPS)      & (\%)        & (\%) & block size & depth\\ 
\hline \hline 
FCFS/R-P  & 3453.91 & 75.3 & 1024 & 4189.17 & 21.29 & 82.3 & 991.23 & 2 \\ \hline
PL/R-P   & 4460.30 & 88.4 & 1024 & 4752.43 & 6.55 & 89.4 & 978.33 & 2 \\ \hline \hline
FCFS/F-P   & 2846.78       & 53.4    & 2048    & 3687.93 & 29.55  & 63.6 & 446.52 & 3 \\ \hline
PL/F-P   & 3381.76 & 68.4 & 2048 & 3614.28 & 6.88  & 66.2 & 1165.70 & 3 \\ \hline \hline
FCFS/EIT-P  & 5650.10 & 91.3 & 1024 & 5747.87 & 1.73  & 92.3 & 1002.26 & 2 \\ \hline
PL/EIT-P & 6096.91 & 93.9 & 1024 & 6206.55 & 1.80  & 95.4 & 1009.91 & 2 \\ \hline \hline
FCFS/EFT-P   & 6581.96 & 23.3 & 2048 & 7569.34 & 15.00 & 63.9 & 412.15  & 5 \\ \hline
PL/EFT-P (*)  &  7046.87 & 55.9 & 2048 & 8030.50 & 13.96 & 86.9 & 407.41 & 4 \\ \hline
\multicolumn{9}{c}{} \\ \hline 
\rowcolor[gray]{.9}
\multicolumn{9}{|c|}{\odroid\ ($8,192\times 8,192$ Cholesky factorization in double precision)} \\ \hline \hline
        & \multicolumn{3}{c||}{\bf Best Homogeneous} & \multicolumn{5}{c|}{\bf Best Found Heterogeneous} \\ \hline \hline
        & Perf.       & Avg. load & Block & Perf.         & Improve.    & Avg. load  & Avg.      & DAG     \\ 
Config. & (GFLOPS)    & (\%)      & size  & (GFLOPS)      & (\%)        & (\%) & block size & depth\\ 
\hline\hline
FCFS/R-P   & 3.75 & 63.9 & 512 & 4.87 & 29.9 & 70.8 & 458.89 & 2 \\ \hline 
PL/R-P     & 4.89 & 70.9 & 512 & 5.84 & 19.3 & 77.4 & 461.11 & 2 \\ \hline \hline
FCFS/F-P   & 7.59 & 69.7 & 512 & 8.10 & 6.74 & 73.7 & 335.80 & 3 \\ \hline 
PL/F-P     & 8.55 & 88.4 & 512 & 8.80 & 2.91 & 92.0 & 466.00 & 2 \\ \hline \hline
FCFS/EIT-P & 8.46 & 98.5 & 256 & 8.52 & 0.76 & 99.1 & 255.19 & 2 \\ \hline 
PL/EIT-P   & 8.74 & 96.2 & 512 & 8.91 & 2.03 & 97.7 & 463.76 & 2 \\ \hline \hline
FCFS/EFT-P   & 8.77 & 89.6 & 512 & 8.96 & 2.20 & 96.2 & 301.23 & 3 \\ \hline 
PL/EFT-P (*) & 8.84 & 91.4 & 512 & 9.08 & 2.75 & 99.0 & 352.07 & 3 \\ \hline

\end{tabular} 
	\end{scriptsize}
\end{table}

\begin{figure}[t]
    \centering
      \begin{subfigure}{\textwidth}
        \includegraphics[width=0.5\textwidth,height=0.7in]{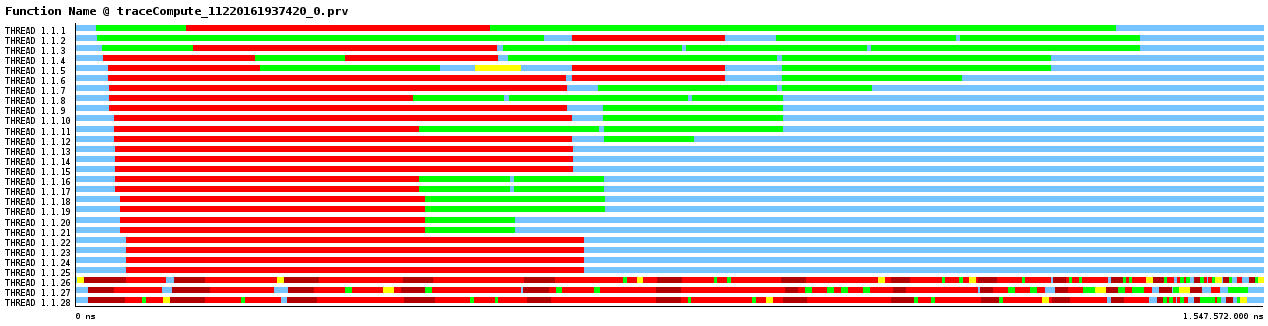} \hspace{0.1cm}
        \includegraphics[width=0.5\textwidth,height=0.7in]{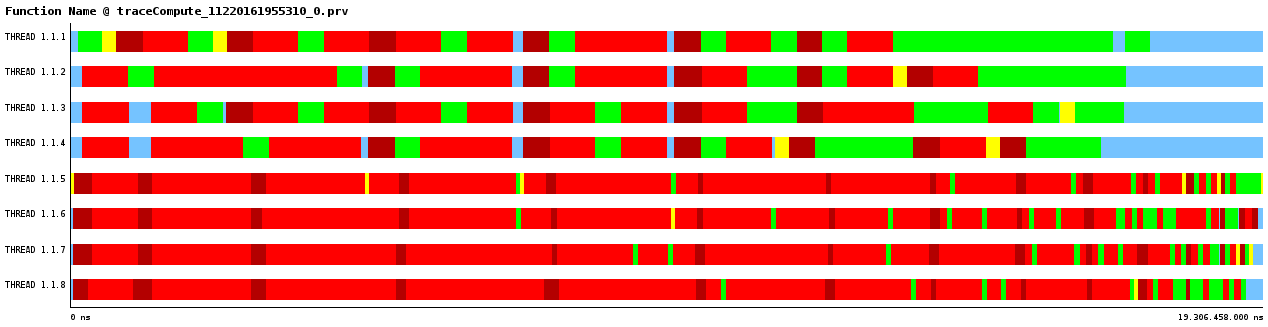}
          \caption{Best homogeneous partitioning. Task scheduling.}
          \label{fig:trace_homo_sched}
      \end{subfigure}

      \begin{subfigure}{\textwidth}
        \vspace{0.2cm}
        \includegraphics[width=0.5\textwidth]{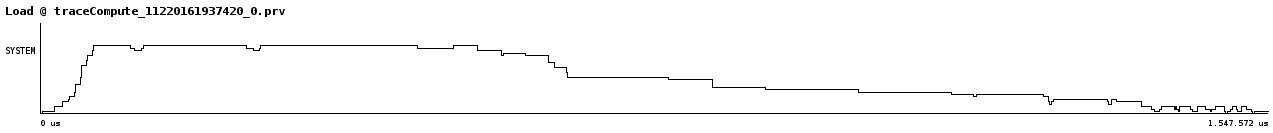} \hspace{0.1cm}
        \includegraphics[width=0.5\textwidth]{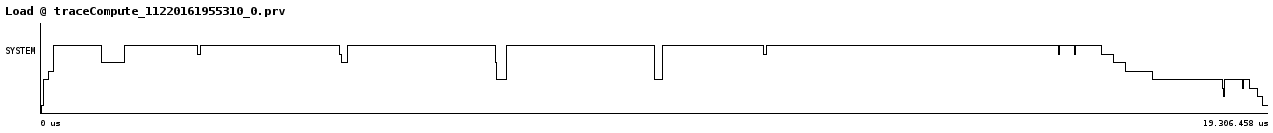}
          \caption{Best homogeneous partitioning. Compute load.}
          \label{fig:trace_homo_load}
      \end{subfigure}

      \begin{subfigure}{\textwidth}
        \vspace{0.2cm}
        \includegraphics[width=0.5\textwidth,height=0.7in]{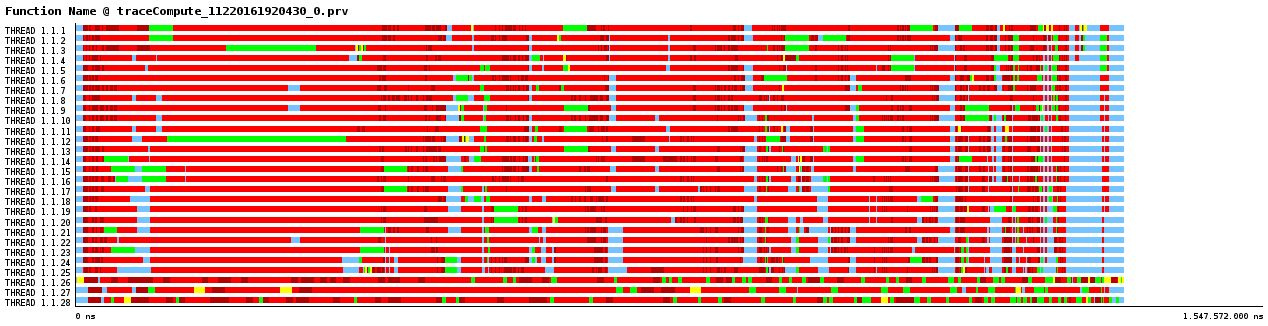} \hspace{0.1cm}
        \includegraphics[width=0.5\textwidth,height=0.7in]{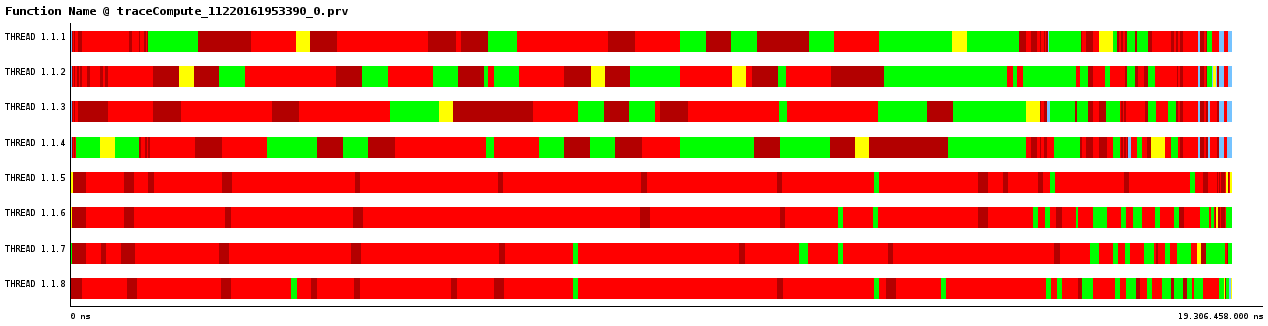}
          \caption{Best heterogeneous partitioning. Task scheduling.}
          \label{fig:trace_hetero_sched}
      \end{subfigure}

      \hfill

      \begin{subfigure}{\textwidth}
        \vspace{0.2cm}
        \includegraphics[width=0.5\textwidth,height=0.7in]{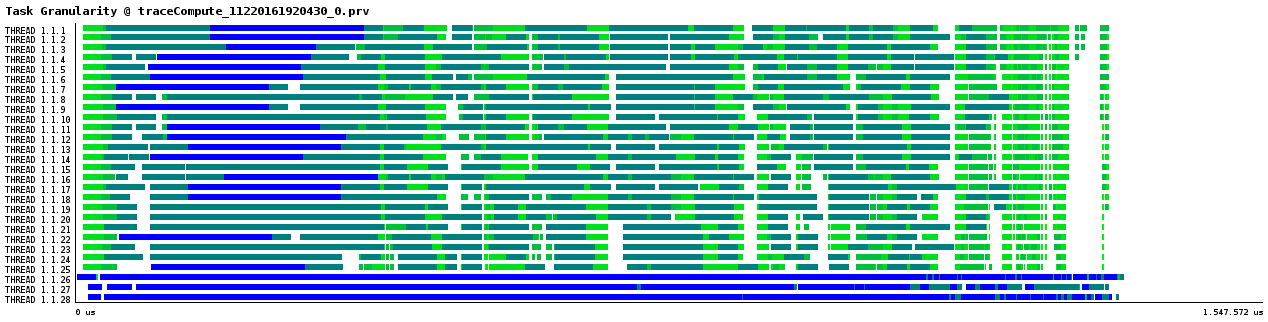} \hspace{0.1cm}
        \includegraphics[width=0.5\textwidth,height=0.7in]{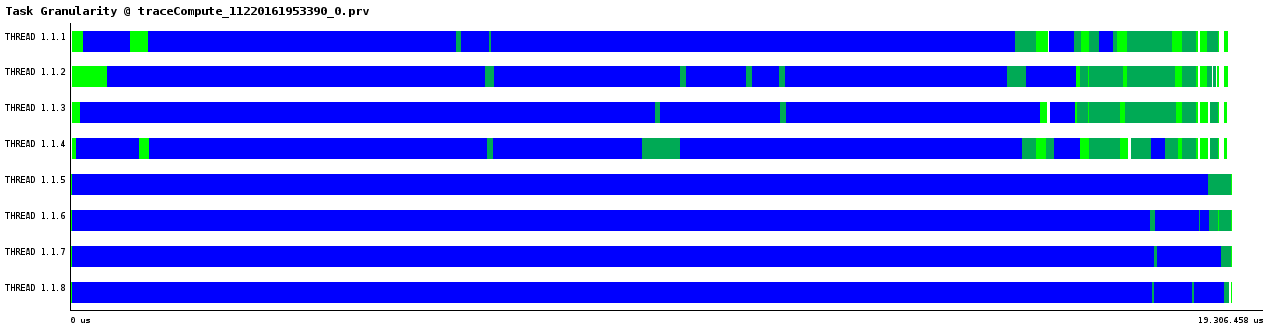}
          \caption{Best heterogeneous partitioning. Task granularity.}
          \label{fig:trace_hetero_granularity}
      \end{subfigure}

      \begin{subfigure}{\textwidth}
        \vspace{0.2cm}
        \includegraphics[width=0.5\textwidth]{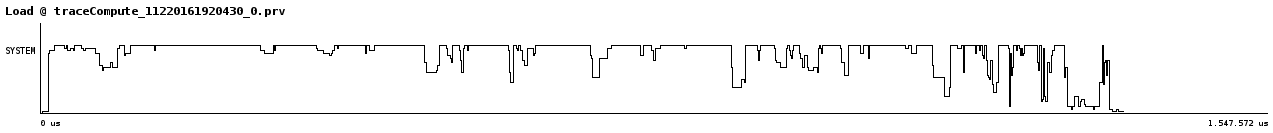} \hspace{0.1cm}
        \includegraphics[width=0.5\textwidth]{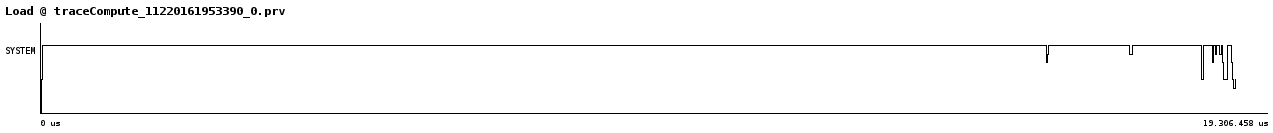}
          \caption{Best heterogeneous partitioning. Compute load.}
          \label{fig:trace_hetero_load}
      \end{subfigure}

\caption{Execution traces for the blocked Cholesky factorization on \bujaruelo\ (left column, $n=32,768$) and \odroid\ (right column, $n=8,192$), using {\sc PL/EFT-P}. For
each case, traces are adjusted to fit the longest execution. In the task scheduling traces, 
colors correspond to the legend in Figure~\ref{fig:taskdata_partitions}.}
\label{fig:traces_ARM}
\end{figure}

Note the direct relation between the average processor occupancy and the improvements of the heterogeneous partitioning scheme. 
For example, EIT-P with homogeneous partitioning yields high processor occupancies (between 91\% and 98.5\%), so the potential benefit expected 
from additional extracted parallelism is poor, ranging between 0.76\% and 2.02\%. Contrary, homogeneous partitions on EFT-P schedules 
yield better performance than EIT-P ones while still leaving more room for potential parallelism. Although the quality of EFT-P schedules could actually 
leave little room for additional improvements, the greater processor availability they offer permits the iterative scheduler-partitioner to find finer-grained 
partitions (see Figure~\ref{fig:trace_hetero_granularity}), attaining remarkable net improvements for \bujaruelo\ (between 13.96\% and 15\%).
Note also that bigger performance improvements do not only correspond with lower processor occupancies, but also with higher task DAG depths (up to 5 in \bujaruelo). This
observation reinforces the importance of managing arbitrary task granularity, introduced by HeSP, extending the idea of using only two degrees of granularity
for two types of processors introduced in other works~\cite{BosilcaHierDAG}.

This reasoning also applies when comparing the highly heterogeneous \bujaruelo\ with the less heterogeneous \odroid\, since the optimal homogeneous tile size seems to fit better to homogeneous platforms, yielding higher occupancies for all scheduling policies tested, hence leaving less room for heterogeneous partitioning improvements. Even with those limitations,
HeSP does always provide improvements in all cases.

Note the even better improvements, with simpler --i.e. less deep-- partitions, attained by our scheme when jointly applied with simpler schedulers --R-P/F-P-- and naive FCFS task ordering. Since bad scheduling decisions exhibit a smaller worsening global impact when applied to a bigger set of smaller tasks, task partitions cooperating with a simple scheduler might alleviate its poor performance: under highly heterogeneous scenarios and available resources, it could be safer to partition a task rather than taking the risk of assigning it to the wrong processor.



Figure~\ref{fig:traces_ARM} reports execution traces for the best-performing configurations observed for both architectures\footnote{Detailed trace generation is
supported by HeSP using Paraver (\url{http://www.bsc.es/computer-sciences/performance-tools/paraver})} (marked in Table~\ref{tab:performance} with an asterisk), using homogeneous and heterogeneous task partitioning
setups. In the traces, each line correspond to a different processor. In \bujaruelo, (25 CPUs on top, 3 GPUs on bottom), observe the amount of idle times (marked in light blue) in the early and
last stages of computation; see how the corresponding optimal heterogeneous schedule fills those gaps by exposing extra parallelism through task partitioning. Concretely,
observing the task granularity trace, in which granularity is reported as a gradient (from light green for small tasks to dark blue for large tasks), it is possible
to conclude that HeSP is able to refine task granularity only on those stages in which processor occupancy is scarce, improving global performance. The increase in compute load can be observed
by comparing the corresponding homogeneous and heterogeneous compute load traces (Figures~\ref{fig:trace_homo_load} and~\ref{fig:trace_hetero_load}).

Similar qualitative results are observed for \odroid, filling the gaps that arise in the same stages on slow cores (top four lines in the trace) 
with finer-grained tasks. In this case, as was observed in Table~\ref{tab:performance}, the opportunities for improvement are more reduced, but overall performance is also increased by our scheme.


%


\section{Conclusion}

In this paper we have presented the HeSP framework and its internal mechanisms towards joint scheduling/partitioning tasks on heterogeneous
architectures. Insights reveal that important performance benefits and improved processor loads can be extracted from the framework for a
family of scheduling policies. The extracted insights for the Cholesky factorization can be easily applied to other irregular task-parallel
implementations, or to arbitrary heterogeneous architectures.

The static iterative implementation of HeSP has shown to be useful to explore the practical performance bounds of a scheduling-partitioning problem, and it naturally paves the road towards a constructive implementation, in which local information is applied on a per-task
basis. This approach can be applied directly on actual task schedulers (e.g. OmpSs) or programming models, in order to introduce in them the recursive task partitioning as an additional degree of freedom.  Future work also includes the exploration of more sophisticated scheduling
techniques attending not only performance optimization, but also energy consumption on different architectures.

%
%
%

\bibliographystyle{plain}
\bibliography{biblio,biblio2,biblio3,biblio4}

\begin{thebibliography}{1}

\bibitem{StarPU}
C{\'e}dric Augonnet, Samuel Thibault, Raymond Namyst, and Pierre-Andr{\'e}
  Wacrenier.
\newblock {StarPU: a unified platform for task scheduling on heterogeneous
  multicore architectures}.
\newblock {\em {CC:PE}}, 23(2):187--198, 2011.

\bibitem{OmpSs}
J.~Bueno, J.~Planas, A.~Duran, R.M. Badia, X.~Martorell, E.~Ayguade, and
  J.~Labarta.
\newblock Productive programming of gpu clusters with ompss.
\newblock In {\em IPDPS 2012}, pages 557--568, May 2012.

\bibitem{Cojean}
Terry Cojean, Abdou Guermouche, Andra Hugo, Raymond Namyst, and
  Pierre-Andr{\'e} Wacrenier.
\newblock {Resource aggregation in task-based applications over
  accelerator-based multicore machines}.
\newblock Technical report, {INRIA}, October 2015.

\bibitem{XKaapi}
Thierry Gautier, Joao~Vicente Ferreira~Lima, Nicolas Maillard, and Bruno
  Raffin.
\newblock {XKaapi: A Runtime System for Data-Flow Task Programming on
  Heterogeneous Architectures}.
\newblock In {\em {27th IPDPS}}, Boston, Massachusetts, United States, May
  2013.

\bibitem{Haidar}
A.~Haidar, A.~YarKhan, Chongxiao Cao, P.~Luszczek, S.~Tomov, and J.~Dongarra.
\newblock Flexible linear algebra development and scheduling with cholesky
  factorization.
\newblock In {\em High Performance Computing and Communications (HPCC), 2015},
  pages 861--864, Aug 2015.

\bibitem{versioning}
J.~Planas, R.M. Badia, E.~Ayguade, and J.~Labarta.
\newblock Self-adaptive ompss tasks in heterogeneous environments.
\newblock In {\em Parallel Distributed Processing (IPDPS), 2013 IEEE 27th
  International Symposium on}, pages 138--149, May 2013.

\bibitem{HEFT}
H.~Topcuoglu, S.~Hariri, and Min-You Wu.
\newblock Performance-effective and low-complexity task scheduling for
  heterogeneous computing.
\newblock {\em Parallel and Distributed Systems, IEEE Transactions on},
  13(3):260--274, Mar 2002.

\bibitem{BosilcaHierDAG}
Wei Wu, A.~Bouteiller, G.~Bosilca, M.~Faverge, and J.~Dongarra.
\newblock Hierarchical dag scheduling for hybrid distributed systems.
\newblock In {\em Parallel and Distributed Processing Symposium (IPDPS), 2015
  IEEE International}, pages 156--165, May 2015.

\end{thebibliography}

\end{document}